\documentclass[aps,prl,twocolumn,showpacs,preprintnumbers,amsmath,amssymb, superscriptaddress]{revtex4}

\usepackage{graphicx}
\usepackage{dcolumn}
\usepackage{bm}

\begin{document} 

\title{Beating the Standard Quantum Limit with Four Entangled Photons}

\author{Tomohisa Nagata}
\affiliation{Research Institute for Electronic Science, Hokkaido University, Sapporo 060-0812, Japan}
\author{Ryo Okamoto}
\affiliation{Research Institute for Electronic Science, Hokkaido University, Sapporo 060-0812, Japan}
\affiliation{Japan Science and Technology Agency, Honcho, Kawaguchi-shi, Saitama, 332-0012, Japan}
\author{Jeremy L. O'Brien}
\affiliation{Department of Electrical and Electronic Engineering, University of Bristol,
Merchant Venturers Building, Woodland Road, Bristol, BS8 1UB, UK}
\affiliation{H. H. Wills Physics Laboratory, University of Bristol, Tyndall Avenue, Bristol BS8 1TL, UK}
\author{Keiji Sasaki}
\affiliation{Research Institute for Electronic Science, Hokkaido University, Sapporo 060-0812, Japan}
\author{Shigeki Takeuchi}
\affiliation{Research Institute for Electronic Science, Hokkaido University, Sapporo 060-0812, Japan}
\affiliation{Japan Science and Technology Agency, Honcho, Kawaguchi-shi, Saitama, 332-0012, Japan}





\begin{abstract}
Precision measurements are important across all fields of science. In particular, optical phase measurements can be used to measure distance, position, displacement, acceleration and optical path length. Quantum entanglement enables higher precision than would otherwise be possible. We demonstrate an optical phase measurement with an entangled four photon interference visibility greater than the threshold to beat the standard quantum limit---the limit attainable without entanglement. These results open the way for new high-precision measurement applications.\\
\\
$^\ast$To whom correspondence should be addressed; E-mail:  takeuchi@es.hokudai.ac.jp
\end{abstract}

\maketitle

Quantum metrology involves using quantum mechanics to realise more precise measurements than can be achieved classically \cite{gi-sci-306-1330}. The canonical example uses entanglement of $N$ particles to measure a phase with a precision $\Delta\phi=1/N$---the Heisenberg limit. Such a measurement outperforms the $\Delta\phi=1/\sqrt{N}$ precision limit possible with $N$ unentangled particles---the standard quantum limit (SQL). Progress has been made with trapped ions \cite{me-prl-86-5870,le-nat--438-639,ro-nat-443-316} and atoms \cite{wi-prl-92-160406,ge-sci-305-270}, while high-precision optical phase measurements have many important applications, including microscopy, gravity wave detection, measurements of material properties, and medical and biological sensing. Although a reduced de Broglie wavelength \cite{ja-prl-74-4835} has been reported for three \cite{mi-nat-429-161}, four \cite{wa-nat-429-158,su-quant-ph/0512212}, and even six \cite{re-quant-ph/0511214} photons, the SQL has been beaten only with two photons \cite{ra-prl-65-1348,ku-qso-10-493,fo-prl-82-2868,ed-prl-89-213601,ei-prl-94-090502}. 

We demonstrate an entangled four photon phase measurement with a visibility that exceeds the threshold to beat the SQL. We use an ultra-stable displaced-Sagnac implementation of a scheme with a high intrinsic efficiency to achieve a four photon interference visibility of 91\%. We also demonstrate that measuring a reduced de Broglie wavelength does not mean beating the SQL, via another experiment which shows high-visibility multi-photon fringes, but can not beat the SQL. The high-precision multi-photon quantum-interference demonstrated here is key, not only to quantum metrology and quantum lithography \cite{fo-prl-82-2868,bo-prl-85-2733,da-prl-87-013602}, but, also to other optical quantum technologies.

The Heisenberg limit and the SQL can be illustrated with reference to an interferometer (Fig. 1, inset) \cite{ho-prl-71-1355,bo-pra-54-4649,ou-pra-55-2598,do-pra-57-4736,ca-pra-68-023810}. We represent a single photon in mode $a$ and no photons in mode $b$ by the quantum state $|10\rangle_{ab}$. After the first beam splitter this photon is in a quantum mechanical superposition of being in both paths of the interferometer: $(|10\rangle_{cd}+|01\rangle_{cd})/\sqrt{2}$. This superposition evolves to the state $(|10\rangle_{cd}+e^{i\phi}|01\rangle_{cd})/\sqrt{2}$ after the $\phi$ phase shift in mode $d$. After recombining at the second beam splitter, the probability of detecting the single photon in mode $e$ is $P_{e}=(1-\cos\phi)/2$, which can be used to estimate $\phi$. If we repeat this experiment $N$ times then the uncertainty in this estimate is $\Delta\phi=1/\sqrt{N}$---the SQL. If instead we were able to prepare the maximally entangled $N$-photon state $(|N0\rangle_{cd}+|0N\rangle_{cd})/\sqrt{2}$ inside the interferometer, this state would evolve to $(|N0\rangle_{cd}+e^{iN\phi}|0N\rangle_{cd})/\sqrt{2}$ after the $\phi$ phase shift. From this state we could estimate the phase with an uncertainty $\Delta\phi=1/N$---the Heisenberg limit---an improvement of $1/\sqrt{N}$ over the SQL. Beating the SQL is known as phase super-sensitivity \cite{mi-nat-429-161,re-quant-ph/0511214}

The $N\phi$ dependence of the phase of the maximally entangled state $(|N0\rangle_{cd}+|0N\rangle_{cd})/\sqrt{2}$ is a manifestation of the $N$-photon de Broglie wavelength $\lambda/N$. This dependence can give rise to an interference oscillation $N$-times faster than that of single photons---phase super resolution \cite{mi-nat-429-161,re-quant-ph/0511214}. Observation of this reduced de Broglie wavelength has sometimes been interpreted in the context of beating the SQL. However, it has been demonstrated recently that high visibility \cite{footnote} $\lambda/N$ resolution can be observed with a purely classical system \cite{re-quant-ph/0511214}. This demonstrates that phase super-resolution by itself does not guarantee a quantum mechanical advantage. Rather, phase super-sensitivity, or beating the SQL, is the most important criterion.

If we put a single photon in each input of the interferometer in Fig. 1(inset), $|11\rangle_{ab}$, the state after the first beam splitter is $(|20\rangle_{cd}+|02\rangle_{cd})/\sqrt{2}$; quantum interference of the two-photon amplitudes cancels the $|11\rangle_{cd}$ term---the Hong-Ou-Mandel (HOM) effect \cite{ho-prl-59-2044}. This state evolves to $(|20\rangle_{cd}+e^{i2\phi}|02\rangle_{cd})/\sqrt{2}$. The probability of detecting two photons in the modes $e$ and $f$, after the second beam splitter, is then $P_{ef}=(1-\cos2\phi)/2$, which shows both phase super-resolution and phase super-sensitivity.

\begin{figure}
\center{\includegraphics[width=1\linewidth]{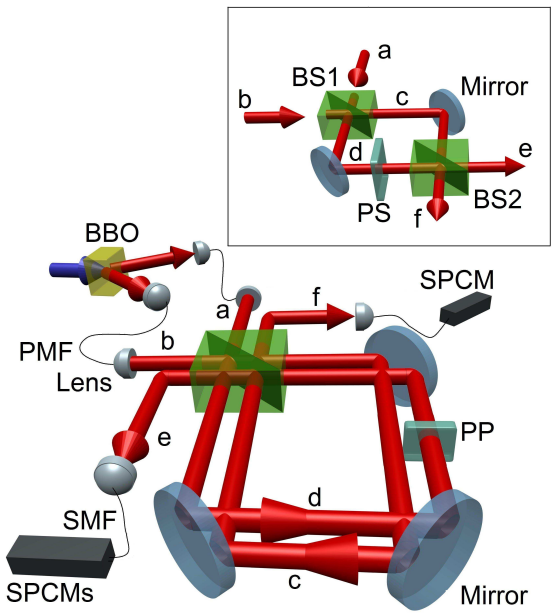}}
\caption{An optical interferometer for beating the SQL. (inset) A schematic of a Mach-Zehnder (MZ) interferometer consisting of two 50:50 beam splitters (BS1 and BS2). Photons are input in modes $a$ and/or $b$, and detected in modes $e$ and/or $f$; after a phase shift (PS) is applied to mode $d$. (main figure)  A schematic of the intrinsically-stable displaced-Sagnac architecture used to ensure that the optical path lengths in modes $c$ and $d$ are sub-wavelength (nm) stable. A frequency doubled 780 nm fs pulsed laser (repetition interval 13 ns) pumps a type-I phase-matched Beta Barium Borate (BBO) crystal to generate the state $|22\rangle_{ab}$ via spontaneous parametric down-conversion. Interference filters (not shown) with a 4 nm bandwidth were used. The photons are guided via polarization maintaining fibres (PMFs) to the interferometer, which has the same function as the MZ interferometer in the inset. A variable phase shift in mode $d$ is realised by changing the angle of the phase plate (PP) in the interferometer. Photons are collected in single mode fibres (SMFs) at the output modes and detected using a single photon counting module (SPCM, detection efficiency 60 \% at 780 nm) in mode $f$ and three cascade SPCMs in mode $e$.}
\end{figure}

Unfortunately this behaviour does not generalise to larger photon number: if we input $|22\rangle_{ab}$, the state after the first beam splitter is:
\begin{equation}
\label{4noon}
\sqrt{\tfrac{3}{4}}(|40\rangle_{cd}+|04\rangle_{cd})/\sqrt{2}+\tfrac{1}{\sqrt{4}}|22\rangle_{cd}
\end{equation}
\noindent
where quantum interference cancels the $|31\rangle_{cd}$ and $|13\rangle_{cd}$ terms \cite{ca-pra-40-1371,ou-prl-83-959}, but the unwanted $|22\rangle_{cd}$ term remains. However, after the second beamsplitter only the $|40\rangle_{cd}$ and $|04\rangle_{cd}$ terms give rise to  $|31\rangle_{ef}$ and $|13\rangle_{ef}$ terms. This is the basis for our experimental scheme \cite{st-pra-65-033820}: The probability of detecting 3 photons in output $e$ and 1 in $f$ is $P_{3ef}=\tfrac{3}{8}(1-\cos4\phi)/2$, which shows phase super-resolution. 

Using state (1) rather than $(|40\rangle_{cd}+|04\rangle_{cd})/\sqrt{2}$, means that our method, like those used previously, can only use a fraction of the initial photons in the $|22\rangle_{ab}$ state, given by the intrinsic efficiency $\eta_i$: $P=\eta_i(1-\cos N\phi)/2$.
To beat the SQL we therefore need to obtain a precision better than $\sqrt{\eta_i/N}$. As the precision of $N$-photon interference with a visibility $V(\le1)$ is $1/VN$, the SQL is beaten for experimentally achieved visibilities above the threshold $V_{th}=1/\sqrt{\eta_i N}$ (cf. \cite{re-quant-ph/0511214}). In our case ($\eta_i=3/8$, $N=4$) $V_{th}=\sqrt{2/3}\approx 81.6\%$. Even though most of the photons pass through the interferometer without leading to a $3ef$ detection event, this scheme can still beat the SQL, since $V_{th}<100\%$. Note that a scheme with $V_{th}>100\%$ can never beat the SQL even with unity efficiency photon sources and detectors.

The existence of $V_{th}$ highlights the need for achieving high-visibility multi-photon interference fringes. Our scheme requires two quantum-inteference and two classical-interference conditions for multi-photon states to be met simultaneously, in a highly time-stable configuration. Therefore we have designed the intrinsically stable displaced-Sagnac experimental architecture shown schematically in Fig. 1.

In order to test the perfomance of our four-photon interferometer, we used pulsed spontaneous parametric down-conversion to produce the four photon input state (see Fig. 1 caption). This source produced not only $|22\rangle$ (2.8 $\times 10^{-4}$ per pulse) but also $|11\rangle_{ab}$ (1.7 $\times 10^{-2}$ per pulse) states. However, $|11\rangle$ states do not contribute to the four-photon coincidence detection. The $|33\rangle$ component (4.7 $\times 10^{-6}$) is two orders of magnitude smaller than $|22\rangle$.

The above discussion requires that the four photon input state be $|22\rangle_{ab}$---four photons in two spatial and one temporal mode---and not $|1111\rangle_{a^{t}a^{t'}b^{t}b^{t'}}$---four photons in two spatial and two temporal modes---\emph{i.e.} the two photons in each mode must be indistinguishable \cite{ei-prl-94-090502,ts-prl-92-153602,ou-pra-72-053814,xi-prl-97-023604}. We have developed a technique for differentiating these two four-photon states using a multi-photon quantum interference generalisation of the HOM effect: if the state $|22\rangle_{ab}$ is input onto a 50:50 beam splitter the probability of detecting two photons in each output is $\tfrac{1}{4}$; while for $|1111\rangle_{a^{t}a^{t'}b^{t}b^{t'}}$ the probability is  $\tfrac{1}{2}$. If there is no quantum interference the probability is $\tfrac{3}{8}$ for both four-photon input states. As we scan the relative delay $\Delta t$ between the arrival time of the photons we move from a regime where $|\Delta t| \gg 0$, and there is no quantum interference, through  $\Delta t=0$, where quantum interference occurs.
From the observed data (Fig. 2), we find that the ratio of the coincidence rate at $\Delta t = 0$ to that at $|\Delta t| \gg 0$ is $\sim \tfrac{2}{3}$, consistent with the ratio of $\tfrac{1}{4}$ to $\tfrac{3}{8}$, indicating that our source generates almost purely $|22\rangle_{ab}$. (Appendix 1.)

\begin{figure}
\center{\includegraphics[width=0.9\linewidth]{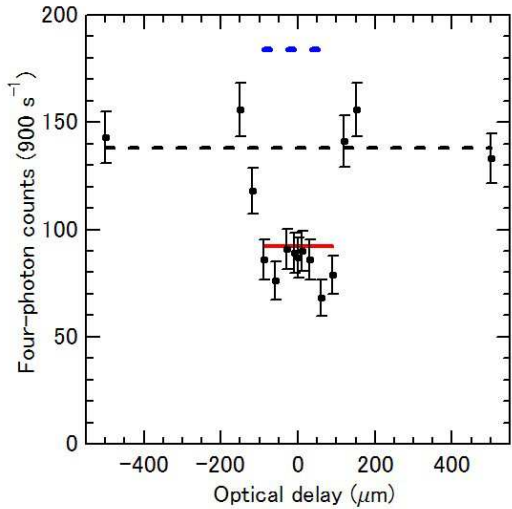}}
\caption{Multi-photon interference at a 50:50 beam splitter demonstrating four-photon indistinguishability. The coincidence count rates of detecting two photons in each output of the beam splitter are recorded as a function of the optical path delay. The ratio of four-photon coincidence counts at delay time $\Delta t=0$ to that at $|\Delta t| \ \gg 0$ is important in order to distinguish the state $|22\rangle_{ab}$ from $|1111\rangle_{a^{t}a^{t'}b^{t}b^{t'}}$ (see text and Appendix 1). The theoretically expected coincidence rate for $|22\rangle_{ab}$ at $ \Delta t \simeq 0$ (red solid line, lower) based on the average count rates at $|\Delta t| \gg 0$ (black dashed horizontal line, middle) agrees well with experimental data (dots). This result indicates that our source almost purely generates $|22\rangle_{ab}$. The dashed blue line (upper) is the theoretically expected coincident rate for $|1111\rangle_{a^{t}a^{t'}b^{t}b^{t'}}$. 
Error bars for this, and the following Figs., show $\pm\sqrt{{\rm counts}}$.}
\end{figure} 

Figure 3 shows the key experimental results of this paper. Firstly we confirm the high visibility classical operation of our interferometer by inputing single photons in mode $a$ $|10\rangle_{ab}$ and detecting the rate of single photons measured in mode $e$; we observe an interference fringe with high visibility ($98\pm0.5\%$) as a function of the optical phase in mode $d$ (Fig. 3A). Next we input pairs of photons $|11\rangle_{ab}$ and measured the rate of two photons detected in modes $e$ and $f$; again we observe a high visibility fringe ($96\pm1\%$), but with half the period of that observed for single photons (Fig. 3B). This visibility is greater than the threshold to beat the SQL ($V_{th}=1/\sqrt{2}$). Finally we input the state  $|22\rangle_{ab}$ and observe the rate of detecting 3 photons in mode $e$ and 1 photon in mode $f$, as described above; again we see a high visibility fringe ($V=91\pm6\%$), now with a period four times smaller than that observed for single photons, demonstrating a four photon de Broglie wavelength (Fig. 3C). More importantly the fringe visibility is greater than the threshold  $V_{th}=\sqrt{2/3}\approx 81.6\%$ to beat the SQL.

\begin{figure}
\center{\includegraphics[width=0.9\linewidth]{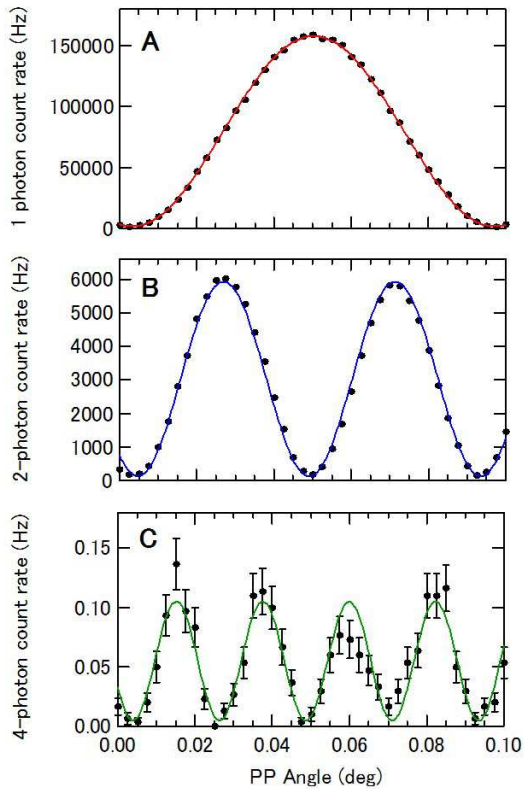}}
\caption{Beating the standard quantum limit with four entangled photons. (\textbf{A}) Single photon count rate in mode $e$ as a function of phase plate (PP) angle with single photon input $|10\rangle_{ab}$. (\textbf{B}) Two photon count rate in modes $e$ and $f$ for input state $|11\rangle_{ab}$. (\textbf{C}) Four photon count rate of 3 photons in mode $e$ and 1 photon in mode $f$ for the input state $|22\rangle_{ab}$. Accumulation times for one data point were (\textbf{A}) 1s, (\textbf{B}) 300s, and (\textbf{C}) 300s.}
\end{figure} 

\begin{figure}
\center{\includegraphics[width=0.9\linewidth]{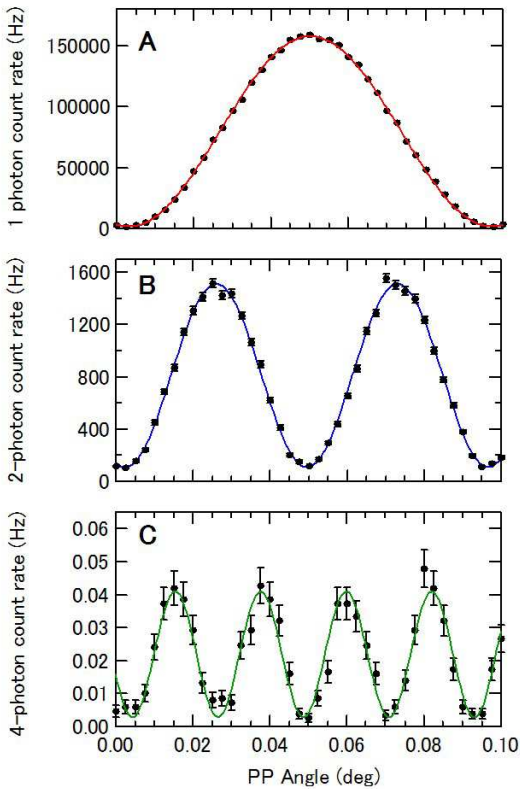}}
\caption{Demonstration of reduced de Broglie wavelength without beating the SQL.  We prepare distinguishable photon inputs by changing the arrival times of the photons at the input to the interferometer and/or changing the detection window. (\textbf{A}) A single photon (classical) interference fringe provides a reference period. (\textbf{B}) Two photons are input into modes $a$ and $b$ at times $t$ and $t'$, where $\delta\tau=t'-t$ is larger than the coherence time of the photons: $|11\rangle_{a^{t}b^{t'}}$. Two photon count rates in mode $e$ are shown with a gate timing window larger than $\delta\tau$. (\textbf{C}) Four photon coincidence events showing phase super-resolution with two independent $|11\rangle_{ab}$ inputs: $|1111\rangle_{a^{t}a^{t'}b^{t}b^{t'}}$. We collected the $|11\rangle_{ef}$ coincidence events which occurred within 200 ns (15 pulses) after a $|20\rangle_{ef}$ event was detected for an accumulation time of 300 s. The count rate per pulse is calculated by including the reduced detection efficiency of the setup. Accumulation times for one data point for (\textbf{A}) and (\textbf{B}) were 1 s.}
\end{figure}

The distinction between a reduced de Broglie wavelength (phase super-resolution) and beating the SQL (phase super-sensitivity) has been described above. An important illustration of this distinction is provided by comparing Figs. 3 and 4; Fig. 4 shows a two photon de Broglie wavelength $\lambda/2$ (B) and a four photon de Broglie wavelength $\lambda/4$ (C), via high visibility ($V=87\pm1\%$ and $V=87\pm5\%$, respectively) interference fringes, but can not demonstrate phase measurement below the SQL. In this case we have used exactly the same experimental setup as before, but have intentionally used distinguishable photon inputs (see Fig. 4 caption). In Fig. 4C we use the input state $|1111\rangle_{a^{t}a^{t'}b^{t}b^{t'}}$, described above. In this case the probability of detecting 3 photons in mode $e$ and 1 in mode $f$ is given by $P_{3ef}=\tfrac{1}{8}(1-\cos4\phi)/2$, which still shows a $4\phi$ dependence on phase. However, because the intinsic efficiency is $\eta_i=\tfrac{1}{8}$, such a scheme can never show a phase sensitivity below the SQL: $V_{th}=\sqrt{2}$. For Fig. 4B the input is $|11\rangle_{a^{t}b^{t'}}$ and the probability of detecting two photons in mode $e$ is $P_{2e}=\tfrac{1}{4}(1-\cos2\phi)/2$, which shows a $2\phi$ phase dependence, but again cannot beat the SQL since $V_{th}=\sqrt{2}$. Note that the reduced count rates in Fig. 4 (B and C) relative to Fig. 3 (B and C) correspond to the reduced $\eta_i$'s

Our results point towards the ultimate sensitivity for optical measurements. To test the performance of the four-photon interferometer, we used a relatively low efficiency source and modest efficiency detectors which means that many more photons pass through the interferometer than lead to a four-photon detection event. For applications (such as biological sensing) where photon flux is important higher efficiency sources might be realized using single photon sources and a HOM interferometer as a heralded two-photon source, while high efficiency number resolving visible light photon counters\cite{ta-apl-99-1063} would dramatically improve detection efficiency. Future possible applications of ultra-high precision phase measurements range from cosmology to medicine.

The authors would like to thank H. F. Hofmann, K. J. Pregnell, G. J. Pryde, J. G. Rarity, K. J. Resch and A. G. White for helpful discussions. This work was supported by the Japan Science and Technology Agency (JST), Ministry of Internal affairs and Communication (MIC), Japan Society for the Promotion of Science (JSPS), 21st century COE program, Special Coordination Funds for Promoting Science and Technology, and the Daiwa Anglo-Japanese Foundation. J.O'B. was supported by an Australian Academy of Science-JSPS Fellowship.

\clearpage

\section*{Appendix 1}
\textit{Testing the indistinguishability of four-photon state.} 
The experiment shown in Figure 2 was performed to confirm that the pair of two photons generated from our source was $|22\rangle_{ab}$ and not $|1111\rangle_{a^t a^{t'} b^t b^{t'}}$. The point here is that the probability of detecting two photons in each output of the beam splitter differs for those two input states. The genuine $|22\rangle_{ab}$ state will show the probability of 1/4, and the state $|1111\rangle_{a^t a^{t'} b^t b^{t'}}$ will show the probability of 1/2. In both cases, a pair of input states ($|2\rangle_a \otimes |2\rangle_b$ or $|11\rangle_{a^t a^{t'}} \otimes |11\rangle_{b^t b^{t'}}$) have to be incident to the beam splitter at exactly the same time, i.e., delay time $\Delta_t = 0$. This is the reason why the probabilities at $\Delta_t = 0$ is essential for distinguishing those two states.

In the experiment, it is more convenient if we have a `standard probability' to which we can compare those probabilities. The probability where no quantum interference occurs among those four photons serves as such a standard. The probability is $3/8$, which can be given by the classical probability to have two out of 4 with 1/2 probability, $_4{\rm C}_2/2^4$.
This situation is realized experimentally by putting two photons in mode a and mode b at different timing, i.e. the path length difference is much more than the spatial distribution of the photonic wave functions (about $150 \mu m$ determined by the filter bandwidth of 4 nm). This provability $3/8$ is the same for both input states $|22\rangle$ and $|1111\rangle$, which is why it serves as a reference standard. This is the reason why the coincidence counts at delay time $|\Delta_t| \gg 0$ is also important.

\vspace{0.2cm}

The reason why the probability is 1/4 for the $|22\rangle_{ab}$ input is easily found in Eq. (1) of the manuscript: The square of the coefficient of the $|22\rangle$ state at the output of the beamsplitter (Eq. (1)) is 1/4.
For $|1111\rangle_{a^t a^{t'} b^t b^{t'}}=|1,1\rangle_{a^t b^t} \otimes |1,1\rangle_{a^{t'} b^{t'}}$ input, the state is converted by a 50:50 beam splitter as follows:

\begin{eqnarray}
&&|1,1\rangle_{\rm a^t,b^t}\otimes|1,1\rangle_{\rm a^{t'},b^{t'}}\rightarrow\nonumber\\
&&\frac{1}{2}\left( |2,0\rangle_{\rm c^t,d^t}+|0,2\rangle_{\rm c^t,d^t}\right)\otimes\left( |2,0\rangle_{\rm c^{t'},d^{t'}}+|0,2\rangle_{\rm c^{t'},d^{t'}}\right) \nonumber
\end{eqnarray}
Among the four possible four output states, the output states where two photons are found in both mode c and mode d are
$|2,0\rangle_{\rm c^t,d^t}\otimes |0,2\rangle_{\rm c^{t'},d^{t'}}$ and $|0,2\rangle_{\rm c^t,d^t}\otimes|2,0\rangle_{\rm c^{t'},d^{t'}}$. Thus, the probability to measure two photons in each output mode c and d is $1/4+1/4=1/2$.

\vspace{0.2cm}

In summary, we can check if the state is $|22\rangle_{ab}$ or $|1111\rangle_{a^t a^{t'} b^t b^{t'}}$ by comparing the four-fold coincidence count rate C($|\Delta_t| = 0$) and C($|\Delta_t| \gg 0$). If C($|\Delta_t| = 0$)/C($|\Delta_t| \gg 0$)=2/3 (0.66), it suggests that the initial state is  $|22\rangle_{ab}$. If C($|\Delta_t| = 0$)/C($|\Delta_t| \gg 0$)=4/3 (1.33), the initial state is $|1111\rangle_{a^t a^{t'} b^t b^{t'}}$. If the output from the source is the mixture of those two states, the ratio is between 2/3 and 4/3. 
It is clear from Fig. 2 that, C($|\Delta_t| = 0$)/C($|\Delta_t| \gg 0$) is almost 0.66, indicating that our source generates almost purely $|22\rangle_{ab}$.


\begin{thebibliography}{10}

\bibitem{gi-sci-306-1330}
V.~Giovannetti, S.~Lloyd, L.~Maccone, {\it Science\/} {\bf 306}, 1330 (2004).

\bibitem{me-prl-86-5870}
V.~Meyer, {\it et~al.\/}, {\it Phys. Rev. Lett.\/} {\bf 86}, 5870 (2001).

\bibitem{le-nat--438-639}
D.~Leibfried, {\it et~al.\/}, {\it Nature\/} {\bf 438}, 639 (2005).

\bibitem{ro-nat-443-316}
C.~F. Roos, M.~Chwalla, K.~Kim, M.~Riebe, R.~Blatt, {\it Nature\/} {\bf 443},
  316 (2006).

\bibitem{wi-prl-92-160406}
A.~Widera, {\it et~al.\/}, {\it Phys. Rev. Lett.\/} {\bf 92}, 160406 (2004).

\bibitem{ge-sci-305-270}
J.~Geremia, J.~K. Stockton, H.~Mabuchi, {\it Science\/} {\bf 304}, 270 (2004).

\bibitem{ja-prl-74-4835}
J.~Jacobson, G.~Bj\"ork, I.~Chuang, Y.~Yamamoto, {\it Phys. Rev. Lett.\/} {\bf
  74}, 4835 (1995).

\bibitem{mi-nat-429-161}
M.~W. Mitchell, J.~S. Lundeen, A.~M. Steinberg, {\it Nature\/} {\bf 429}, 161
  (2004).

\bibitem{wa-nat-429-158}
P.~Walther, {\it et~al.\/}, {\it Nature\/} {\bf 429}, 158 (2004).

\bibitem{su-quant-ph/0512212}
F.~W. Sun, B.~H. Liu, Y.~F. Huang, Z.~Y. Ou, G.~C. Guo, {\it
  quant-ph/0512212\/}  (2005).

\bibitem{re-quant-ph/0511214}
K.~J. Resch, {\it et~al.\/}, {\it quant-ph/0511214\/}  (2005).

\bibitem{ra-prl-65-1348}
J.~G. Rarity, {\it et~al.\/}, {\it Phys. Rev. Lett.\/} {\bf 65}, 1348 (1990).

\bibitem{ku-qso-10-493}
A.~Kuzmich, L.~Mandel, {\it Quant. Semiclass. Opt.\/} {\bf 10}, 493 (1998).

\bibitem{fo-prl-82-2868}
E.~J.~S. Fonseca, C.~H. Monken, S.~P\'adua, {\it Phys. Rev. Lett.\/} {\bf 82},
  2868 (1999).

\bibitem{ed-prl-89-213601}
K.~Edamatsu, R.~Shimizu, T.~Itoh, {\it Phys. Rev. Lett.\/} {\bf 89}, 213601
  (2002).

\bibitem{ei-prl-94-090502}
H.~S. Eisenberg, J.~F. Hodelin, G.~Khoury, D.~Bouwmeester, {\it Phys. Rev.
  Lett.\/} {\bf 94}, 090502 (2005).

\bibitem{bo-prl-85-2733}
A.~N. Boto, {\it et~al.\/}, {\it Phys. Rev. Lett.\/} {\bf 85}, 2733 (2000).

\bibitem{da-prl-87-013602}
M.~D'Angelo, M.~V. Chekhova, Y.~Shih, {\it Phys. Rev. Lett.\/} {\bf 87}, 013602
  (2001).

\bibitem{ho-prl-71-1355}
M.~J. Holland, K.~Burnett, {\it Phys. Rev. Lett.\/} {\bf 71}, 1355 (1993).

\bibitem{bo-pra-54-4649}
J.~J. Bollinger, W.~M. Itano, D.~J. Wineland, D.~J. Heinzen, {\it Phys. Rev.
  A\/} {\bf 54}, 4649 (1996).

\bibitem{ou-pra-55-2598}
Z.~Y. Ou, {\it Phys. Rev. A\/} {\bf 55}, 2598 (1997).

\bibitem{do-pra-57-4736}
J.~P. Dowling, {\it Phys. Rev. A\/} {\bf 57}, 4736 (1998).

\bibitem{ca-pra-68-023810}
R.~A. Campos, C.~C. Gerry, A.~Benmoussa, {\it Phys. Rev. A\/} {\bf 68}, 023810
  (2003).

\bibitem{footnote}
The visibility $V\equiv(max-min)/(max+min)$, where $max$ and $min$ are the
  extreme values of the sinusoidal multi-photon interference fringe; $0\le V\le
  1$.

\bibitem{ho-prl-59-2044}
C.~K. Hong, Z.~Y. Ou, L.~Mandel, {\it Phys. Rev. Lett.\/} {\bf 59}, 2044
  (1987).

\bibitem{ca-pra-40-1371}
R.~A. Campos, B.~E.~A. Saleh, M.~C. Teich, {\it Phys. Rev. A\/} {\bf 40}, 1371
  (1989).

\bibitem{ou-prl-83-959}
Z.~Y. Ou, J.-K. Rhee, L.~J. Wang, {\it Phys. Rev. Lett.\/} {\bf 83}, 959
  (1999).

\bibitem{st-pra-65-033820}
O.~Steuernagel, {\it Phys. Rev. A\/} {\bf 65}, 033820 (2002).

\bibitem{ts-prl-92-153602}
K.~Tsujino, H.~F. Hofmann, S.~Takeuchi, K.~Sasaki, {\it Phys. Rev. Lett.\/}
  {\bf 92}, 153602 (2004).

\bibitem{ou-pra-72-053814}
Z.~Y. Ou, {\it Phys. Rev. A\/} {\bf 72}, 053814 (2005).

\bibitem{xi-prl-97-023604}
G.~Y. Xiang, {\it et~al.\/}, {\it Phys. Rev. Lett.\/} {\bf 97}, 023604 (2006).

\bibitem{ta-apl-99-1063}
S.~Takeuchi, J.~Kim, Y.~Yamamoto, and H.~H. Hogue, {\it Appl. Phys. Lett.\/} {\bf 74}, 1063 (1999).

\end{thebibliography}
\end{document}